\documentclass[%
article,
twocolumn,
superscriptaddress,
amsmath,amssymb,
aps,
journal=apl,
]{revtex4-1}

\usepackage{natbib}
\usepackage{graphicx}
\usepackage{dcolumn}
\usepackage{bm}
\usepackage{xcolor}
\usepackage{float}
\usepackage{newtxtext,newtxmath}

\begin{document}

\title{
Simulating anharmonic vibrational polaritons beyond the long wavelength approximation
}

\author{Dipti Jasrasaria}
\email{dj2667@columbia.edu}
\affiliation{Department of Chemistry, Columbia University, New York, New York 10027, USA}

\author{Arkajit Mandal}
\email{mandal@tamu.edu}
\affiliation{Department of Chemistry, Columbia University, New York, New York 10027, USA}
\affiliation{Department of Chemistry, Texas A\&M University, College Station, Texas 77843, USA}

\author{David R. Reichman}
\email{drr2103@columbia.edu}
\affiliation{Department of Chemistry, Columbia University, New York, New York 10027, USA}

\author{Timothy C. Berkelbach}
\email{t.berkelbach@columbia.edu}
\affiliation{Department of Chemistry, Columbia University, New York, New York 10027, USA}
\affiliation{Initiative for Computational Catalysis, Flatiron Institute, New York, New York 10010, USA}

\begin{abstract}
 In this work we investigate anharmonic vibrational polaritons formed due to strong light-matter interactions in an optical cavity between radiation modes and anharmonic vibrations beyond the long-wavelength limit. We introduce a conceptually simple description of light-matter interactions, where spatially localized cavity radiation modes couple to localized vibrations. Within this theoretical framework, we employ self-consistent phonon theory and vibrational dynamical mean-field theory to efficiently simulate momentum-resolved vibrational-polariton spectra, including effects of anharmonicity. Numerical simulations in model systems demonstrate the accuracy and applicability of our approach. 
\end{abstract}

\date{\today}
\maketitle

\section{Introduction}

Strong coupling between light and matter inside optical cavities lead to the formation of polaritons, light-matter hybrid quasiparticles that exhibits a wide variety of exotic physical and chemical effects~\cite{ThomasS2019, MandalCR2023, BhuyanCR2023, XuNC2023, SimpkinsCR2023, HiraiCR2023, BasovNp2021, Hutchison2012ACIE, Feist2018AP, Herrera2016PRL, GarciaVidal2021S, Kowalewski2017PNASU, Dunkelberger2022ARPC, Barquilla2022ACSP, Kockum2019NRP, RuggenthalerCR2023}. Examples include the possible modification of both ground~\cite{ThomasS2019, Hutchison2012ACIE, Wonmi2023S, AnoopNp2020, Nagarajan2021JACS} and excited state chemical reactivities~\cite{InkiJACS2024, ZengJACS2023, Hutchison2012ACIE}, enhanced exciton~\cite{XuNC2023, BalasubrahmaniyamNM2023,BerghuisACSP2022, PandyaAS2022} and charge carrier~\cite{NagarajanACSN2020} transport, modification of crystallization and melting processes~\cite{HiraiCS2021, Brawley_Yim_Pannir-Sivajothi_Poh_Yuen-Zhou_Sheldon_2023}, and long-range exciton energy transfer~\cite{Son2022}. However, many of these effects remain elusive experimentally due to a lack of clear theoretical understanding despite the significant effort and progress in recent years~\cite{LiNC2021, LindoyJPCL2022, lindoy2023quantum, YingCM2024, AnguloJCP2023, LiARPC2022,WangACSP2021, mondal2022dissociation, SchaferNC2022}. 

Currently, many theoretical and experimental works operate in mutually incompatible parameter regimes, resulting in contradictory observations~\cite{MandalCR2023, BhuyanCR2023, AnguloJCP2023, LiARPC2022}. While many theoretical studies employ a single emitter and a single cavity mode description~\cite{LiNC2021, LindoyJPCL2022, YingCM2024, SchaferNC2022, WeightJACS2024, GalegoNC2016, PavosevicNC2023, SunJPCL2023}, a majority of experiments operate in the collective regime, where an ensemble of molecules are coupled to an ensemble of cavity radiation modes~\cite{Wonmi2023S,
Hutchison2012ACIE, AnoopNp2020, ZengJACS2023, ThomasS2019}. Recent works have demonstrated that a multi-mode-multi-molecule description (i.e., beyond long-wavelength approximation) is necessary to capture various experimentally-observed photo-physical properties, such as cavity modified exciton transport~\cite{XuNC2023, BerghuisACSP2022, Sokolovskii2023NC, Ribeiro2024Np}, polariton relaxation and thermalization~\cite{TichauerJCP2021, FreireACSN2024}, polariton lasing~\cite{ArnardottirPRL2020}, polariton condensation~\cite{KeelingARPC2020}, angle resolved polariton  spectra~\cite{QiuJPCL2021, MandalNL2023}, and polaritonic up-conversion~\cite{MiteshACSN2024}. Thus, describing polariton systems beyond the long wavelength approximation may be relevant in various cavity modified chemical and physical effects in molecules and materials and may be key to resolving discrepancies between theoretical and experimental work.


Despite notable progress in multiscale polaritonic simulations~\cite{LukJCTC2017, li2024vibrational, TichauerJCP2021}, simulating a large ensemble of cavity radiation modes and molecules remains a computationally formidable task. New theoretical frameworks may offer opportunities to develop computationally and conceptually convenient approaches. In this work, we develop a theoretical framework where light-matter interactions occur between localized cavity radiation modes and vibrations. While this description is formally equivalent to a dipole-gauge Hamiltonian beyond the long wavelength approximation, it allows for more efficient spatial truncation and spatial coarse graining of the light-matter hybrid systems. Using this new description, we employ vibrational dynamical mean-field theory (VDMFT)~\cite{shih_anharmonic_2022, jasrasaria2024} to study anharmonic phonon-polaritons formed by coupling the vibrations of a periodic lattice to quantized radiation modes inside an optical cavity. Within this approach the spectra of an extended light-matter hybrid system is simulated via an impurity model, which maps the dynamics of the periodic system to that of a single unit cell coupled to a self-consistently defined bath of harmonic oscillators. It is worth noting that the typical dipole-gauge Hamiltonian beyond the long wavelength approximation, where cavity radiation modes are spatially delocalized, as has been used recently to simulate vibrational polaritons~\cite{li2024vibrational}, is incompatible with VDMFT, which requires spatially localized description, as offered within our new description of light and matter.

Here, we benchmark our approach in a simple model molecular system coupled to cavity radiation modes. Using our approach, we compute the momentum-resolved phonon-polariton spectra and find it to be very accurate in comparison to the exact spectra computed using molecular dynamics simulations. We show that effects of anharmonicity in the molecular system, which are often ignored, play a crucial role determining both energies and linewidths in the momentum-resolved phonon-polariton spectral function, which have implications for heat transport and other phenomena, and that harmonic approximations often employed to fit experimental data can break down. We also find that the presence of anharmonicity leads to a nonlinear relationship between the Rabi splitting and light-matter coupling strength in contrast to the linear relationship predicted in a simple coupled harmonic oscillator model. Our work also illustrates the shortcoming of simple light-matter models to extract light-matter couplings in experiments and underscores the importance of accurately modeling molecular and photonic degrees of freedom as well as their interactions. 

The rest of this paper is structured as follows. In Section~\ref{sec:model}, we discuss the model light-matter hybrid system used in our work. In Section~\ref{sec:hamiltonian}, we introduce a new light-matter Hamiltonian that describes spatially localized photons interacting with matter. In Section~\ref{sec:dynamics} we briefly describe self-consistent phonon theory and vibrational dynamical mean-field theory. In Section~\ref{sec:results}, we present our numerical results and discuss their implications. Finally, in Section~\ref{sec:conclusions}, we summarize our work and document our conclusions. 

\section{Model\label{sec:model}}

We consider a one-dimensional periodic chain of atoms in a cavity that is perpendicular to the confined direction of the cavity. In this work, for simplicity, we consider a 2D world where cavity confinement is along the $x$ direction and two mirrors are placed along the $y$ direction. A cavity radiation mode has momentum $\bm{k} = k_x \bm{\hat{x}} + k_y \bm{\hat{y}}$ and photon frequency $\omega_c(\bm{k})=c\sqrt{k_x^2 +k_y^2}$ (we set the refractive index $\eta_r = 1$), where $k_x=n\pi/L_x$ and $k_y=2n'\pi/L_y$ with $n,n'$ as positive integers. Here, $L_x$ is the distance between the two mirrors, and we use periodic boundary conditions along the $\hat{\bm y}$ direction with a supercell length of $L_y$. Due to the relevant energy scales, we consider only $n=1$, such that $k_x=\pi/L_x\equiv\omega_0/c$ and $\omega_c(k_y)=\sqrt{\omega_0^2 + c^2 k_y^2}$. Further, we only consider the transverse electric polarization, although the analysis would be similar if we considered the transverse magnetic polarization.

We consider a light-matter Hamiltonian beyond the long-wavelength approximation~\cite{MandalCR2023, MandalNL2023, DmytrukPRB2021}, which is given by (using atomic units $\hbar = 1$ and mass-weighted coordinates)
%
\begin{align}
    H &= H_L + H_M + H_{L-M} + H_{DSE} \label{eqn:fullH} \\
    H_L &= \sum_{k} \omega_c(k) \big(\hat{a}_k^\dagger \hat{a}_k + \frac{1}{2}\big) \\
    H_M &= \frac{1}{2} \sum_j \big( \dot{r}_j^2 + \omega_m^2 r_j^2 + gr_j^4 + \Omega_m^2 (r_j - r_{j+1})^2 \big) \label{eqn:H_M} \\
    H_{L-M} &= \sum_k \sum_j \eta \sqrt{2 \omega_0 \omega_m \omega_c(k)} \cos\theta_k \nonumber \\
    & \qquad \qquad \times \big(\hat{a}_k^\dagger e^{-ikR_j} + \hat{a}_k e^{ikR_j}\big) r_j \\
    H_{DSE} &= \sum_{k}\sum_{j,j'}2 \eta^2 \omega_0 \omega_m \cos^2 \theta_k e^{ik(R_j - R_{j'})} r_j r_{j'} \,.
\end{align}
This Hamiltonian was derived by assuming that the electromagnetic field varies slowly over the single unit cell~\cite{JiajunPRB2020, MandalNL2023, DmytrukPRB2021} and is consistent with the form of the generalized Tavis-Cummings Hamiltonian~\cite{DunkelbergerARPC2022, MandalCR2023}. Here, $H_L$ is the Hamiltonian that describes the cavity, where $k\equiv k_y$, and $\hat{a}_k^\dagger$ and $\hat{a}_k$ are the creation and annihilation operators, respectively, of a cavity mode with momentum $k$. Further, $\theta_k$ is the angle between the polarization of cavity mode $k$ and matter dipoles oriented in the $y$ direction. 

$H_M$ is the matter Hamiltonian for the isolated chain of atoms and, in principle, can take any form. Here, we use a model of coupled, local oscillators, where $j$ is an index over lattice sites, and $r_j = \sqrt{\frac{1}{2\omega_m}}(\hat{b}^\dagger_j + \hat{b}_j)$ is the displacement from equilibrium of an atom at site $j$, and $\hat{b}_j^\dagger$ and $\hat{b}_j$ are the vibrational creation and annihilation operators, respectively, of the atomic vibration at site $j$. 

The light-matter coupling is given by $H_{L-M}$, where $\eta$ indicates the coupling strength between the field and the matter, $\eta\equiv 1/\sqrt{\varepsilon_0 \omega_0 V}$, where $V$ is the cavity volume, and $R_j$ is the equilibrium position of atom at site $j$. $H_{DSE}$ describes the dipole self-energy.

\begin{figure}[!ht]
    \centering
    \includegraphics[width=8.5cm]{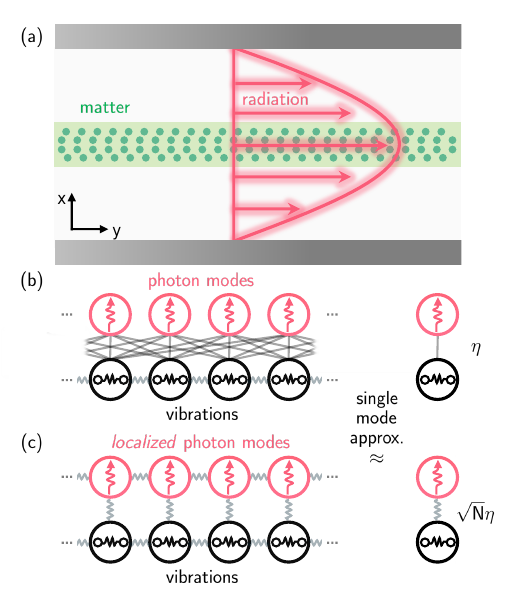}
    \caption{
    (a) Schematic representation of a material coupled to a quantized radiation mode inside an optical cavity. Schematic representations of a lattice of vibration modes coupled to a set of cavity radiation modes within (b) the dipole-gauge Hamiltonian beyond the long-wavelength approximation and (c) the real-space light-matter Hamiltonian introduced in this work. The single cavity mode - single matter vibration limits are illustrated on the right.}
    \label{fig1:schematic}
\end{figure}

Note that this Hamiltonian goes beyond the long wavelength approximation, as noted by the complex phase factor, $e^{i k R_j}$, which describes the spatial variation of the radiation. 

\section{Transforming to real-space\label{sec:hamiltonian}}

We transform the Hamiltonian given by Eq.~(\ref{eqn:fullH}) to real-space using canonical transformations between cavity mode operators, $x_k = \sqrt{\frac{1}{2\omega_c(k)}}(\hat{a}^\dagger_{-k} + \hat{a}_k)$, and unitary transformations between real-space and Fourier-space, $x_j = \sqrt{\frac{1}{N}} \sum_k e^{ikR_j} x_k$.

\subsection{Cavity Hamiltonian \label{subsec:cavityHamiltonian}}


We start by transforming the Hamiltonian of the isolated cavity:
\begin{align}
    H_L &= \sum_{k} \omega_c(k) \big(\hat{a}_k^\dagger \hat{a}_k + \frac{1}{2}\big) \\
    &= \sum_{j} \frac{1}{2} \big( \dot{x}_j^2 + \omega_0^2x_{j}^2\big)  + \frac{c^2}{2}\sum_k k^2x_{-k}x_{k}\,.
\end{align}

Now, we will focus on the final term of the above equation, whose Fourier transform we will take by rewriting the expression as a derivative approximated using a second-order central finite difference:
\begin{align}
    &\sum_k k^2x_{-k}x_{k} = \frac{1}{N} \sum_k \sum_{j,j'} k^2 e^{ik(R_j - R_{j'})} x_j x_{j'} \nonumber \\
    &= -\frac{1}{N} \sum_k \sum_{j,\delta j} \frac{\partial^2 e^{ika\delta j}}{\partial(\delta j)^2} x_j x_{j-\delta j} \label{eqn:2ndDerivative}\nonumber \\
    &\approx -\frac{1}{a^2} \sum_j \big( x_j x_{j+1} - 2x_j^2 + x_j x_{j-1}\big)\,,
\end{align}
where $a$ is the lattice constant (i.e., the distance between adjacent lattice sites), and $\delta j = j - j'$.

Thus, the full cavity Hamiltonian in real-space is given by
\begin{equation}
    H_L = \frac{1}{2}\sum_j \big( \dot{x}^2_j + \omega_0^2 x_j^2 + \frac{c^2}{a^2}(x_j - x_{j+1})^2 \big)\,,
\end{equation}
indicating that the isolated cavity mode can be described using a one-dimensional chain of ``cavity" atoms with localized, harmonic vibrations and nearest-neighbor harmonic interactions. Note that the second derivative in Eq.~(\ref{eqn:2ndDerivative}) can be approximated using higher-order central finite difference, which would introduce longer range harmonic interactions between cavity atoms, but this higher-order expansion is not necessary, as described further below in Sec.~\ref{sec:results}.

\subsection{Light-matter coupling}

Next, we address the light-matter coupling:
\begin{align}
    H_{L-M} 
    &= \sum_k \sum_j \eta \sqrt{2 \omega_0 \omega_m \omega_c(k)} \cos\theta_k \nonumber \\
    &\qquad \qquad \times \big(\hat{a}_k^\dagger e^{-ikR_j} + \hat{a}_k e^{ikR_j}\big) r_j \\
    &= \sum_j 2 \eta \sqrt{N \omega_0^3 \omega_m} x_j r_j\,,\label{eqn:l-m-sqrtN}
\end{align}
where we have made use of the identity $\cos\theta_k = \omega_0/\omega_c(k)$.

\subsection{Dipole self-energy}

Finally, we can transform the dipole self-energy term:
\begin{align}
    H_{DSE} 
    &= \sum_{k}\sum_{j,j'} 2\eta^2 \omega_0 \omega_m \cos^2 \theta_k e^{ik(R_j - R_{j'})} r_j r_{j'} \\
    &\approx \sum_{j} 2\eta^2 \omega_0 \omega_m r_j^2\,,
\end{align}
where we have assumed $\omega_0/\omega_c^2(k) \approx 1$, which is true for $ck_y \ll \omega_0$.

Thus, we have transformed the light-matter Hamiltonian into a purely real-space Hamiltonian that mirrors that of a one-dimensional lattice, where each unit cell consists of two atoms --- one ``cavity" atom (a localized cavity radiation mode) and one ``matter" atom --- that are bilinearly coupled to each other, and like atoms in adjacent unit cells are coupled to one another via harmonic interactions:
\begin{align} 
    H = &\frac{1}{2}\sum_j \big( \dot{x}^2_j + \omega_0^2 x_j^2 + \frac{c^2}{a^2}(x_j - x_{j+1})^2 \big) + H_M \nonumber \\
    &+ \sum_j 2 \eta \sqrt{N \omega_0^3 \omega_m} x_j r_j + \sum_{j} 2\eta^2 \omega_0 \omega_m r_j^2\,.
    \label{eqn:H_polaritonLattice}
\end{align}

The structures of the light-matter couplings in the original dipole gauge and our real-space picture are illustrated in Fig.~\ref{fig1:schematic}. It is important to note that unlike the dipole gauge Hamiltonian, where light-matter couplings do not scale with the number of matter degrees of freedom $N$,  in  Eq.~(\ref{eqn:H_polaritonLattice}) the light-matter coupling scales with $\sqrt{N}$. 
This provides an enticing perspective on a fundamental question in polariton chemistry~\cite{LindoyNp2024}: 

\emph{Can collective light–matter coupling, which couples cavity radiation and molecular degrees of freedom (DOF) in a delocalized fashion, lead to a modification of chemical reactivity that operates locally?} 

This question above is posed in the context of reducing the dipole-gauge Hamiltonian in Eq.~(\ref{eqn:fullH}) to a single molecular and photonic degree of freedom. The consequence of such an approximation (illustrated in Fig.~\ref{fig1:schematic}b) is that the light-matter interaction term reduces to
\begin{align}
    &\sum_k \sum_j \eta \sqrt{2 \omega_0 \omega_m \omega_c(k)} \cos\theta_k \big(\hat{a}_k^\dagger e^{-ikR_j} + \hat{a}_k e^{ikR_j}\big) r_j \nonumber \\
    &\qquad \rightarrow \eta \omega_0 \sqrt{2  \omega_m } \cos\theta_0  \big(\hat{a}^\dagger   + \hat{a}  \big) r_0\,,
\end{align}
such that the coupling between a single molecule and a single cavity mode is weaker by a factor of $1/\sqrt{N}$ when comparing to the collective Rabi-splitting. In contrast, when the same approximation is made in the real-space light-matter Hamiltonian, that is
\begin{equation}
    \sum_j 2 \eta \sqrt{N \omega_0^3 \omega_m} x_j r_j \rightarrow \eta \sqrt{N \omega_0^3 \omega_m} x_0 r_0\,,
\end{equation}
the single molecule-single cavity mode coupling scales as $\sqrt{N}$, as illustrated in Fig.~\ref{fig1:schematic}c. Note that $\eta \propto 1/\sqrt{V}$ such that the single molecule-localized cavity mode coupling is proportional to the density $\sqrt{N/V}$ (i.e., concentration) of matter DOF inside the optical cavity. Here, $N$ represents the number of molecular DOF placed in the plane of the mirrors and that the number molecular DOF in the direction perpendicular to the mirrors' plane would lead to a dilution of the light-matter coupling. 

Overall, beyond its computational utility, the theoretical description introduced here opens up new  questions regarding cavity-modified chemical phenomena. To what extent these localized cavity modes modify chemical reactivity locally also remains an open question. 

\section{Modeling lattice (polariton) dynamics\label{sec:dynamics}}

We characterize the polariton system by calculating its lattice dynamics according to different levels of theory.

\subsection{Harmonic dynamics}


A harmonic description of the polariton lattice dynamics is given by the dynamical matrix,
\begin{equation}
    \mathcal{D}_{\alpha,\alpha'}(k) = \sum_j e^{ik(R_{j\alpha} - R_{j'\alpha'})} \frac{\partial^2 \mathcal{V}}{\partial u_{j\alpha} \partial u_{j'\alpha'}}\,,
\end{equation}
where $k$ is a wavevector in the first Brillouin zone (BZ), $\alpha,\alpha'$ index the atoms at each lattice site (i.e., either the cavity or the matter atom), $R_{j\alpha}$ is the equilibrium position of atom $\alpha$ at lattice site $j$, and $u_{j\alpha}$ is the displacement away from the equilibrium position. Derivatives of the lattice potential, $\mathcal{V}$, are evaluated at the equilibrium lattice configuration.

Diagonalization of the dynamical matrix yields the phonon modes and frequencies as the eigenvectors and the square root of the eigenvalues, respectively. Thus, the coordinate of phonon mode $\lambda$ is given  by
\begin{equation}
    u_\lambda(k) = N^{-1/2} \sum_{j\alpha} c_{\alpha,\lambda}(k) e^{-ikR_{j\alpha}} u_{j\alpha}\,,
\end{equation}
where $\bm{c}(k)$ are the eigenvectors of $\bm{\mathcal{D}}(k)$.

\subsection{Self-consistent phonon theory}

The harmonic picture can be improved upon using self-consistent phonon (SCP) theory~\cite{Hooton1958, KoehlerPRL1966, werthamer1970self, klein1972rise,tadano2018first}, which treats anharmonicity at a static mean-field theory level. Temperature-dependent anharmonic frequencies and eigenvectors are calculated by self-consistently solving the equation
\begin{align}
    V_{\lambda,\lambda'}(\bm{k}) = &\omega_\lambda^2(\bm{k})\delta_{\lambda,\lambda'} \nonumber \\
    &+ \frac{1}{2}\sum_{\bm{k}'}\sum_{\lambda'',\lambda'''} \Phi_{\lambda,\lambda',\lambda'',\lambda'''}(\bm{k},-\bm{k},\bm{k}',-\bm{k}') \nonumber \\
    &\qquad \qquad \qquad \times \langle Q_{\lambda''}^* (\bm{k}') Q_{\lambda'''}(\bm{k}')\rangle\,,
\end{align}
where $\omega_\lambda(\bm{k})$ is the harmonic frequency of mode $\lambda$ at $\bm{k}$, $\Phi_{\lambda,\lambda',\lambda'',\lambda'''}(\bm{k},-\bm{k},\bm{k}',-\bm{k}')$ is the reciprocal representation of the fourth-order interatomic force constants computed using the harmonic eigenvectors of the dynamical matrix, and
\begin{equation}
    \langle Q_{\lambda''}^* (\bm{k}') Q_{\lambda'''}(\bm{k}')\rangle = \sum_\mu U_{\lambda'',\mu}(\bm{k}') \frac{k_BT}{\Omega^2_\mu(\bm{k}')} U_{\mu,\lambda'''}(\bm{k}')\,,
\end{equation}
where $\bm{U}(\bm{k})$ are the eigenvectors of $\bm{V}$ (i.e., the unitary matrix that transforms the harmonic phonon eigenvectors into the anharmonic ones), and $\Omega_\mu(\bm{k})$ is the renormalized frequency of anharmonic mode $\mu$ at $\bm{k}$. The above equation assumes the high-temperature limit of classical statistics. In this simple model the fourth-order force constant simplifies to
\begin{align}
    &\Phi_{\lambda,\lambda',\lambda'',\lambda'''}(\bm{k},-\bm{k},\bm{k}',-\bm{k}') = \nonumber \\
    & \qquad \qquad N^{-1}c_{m,\lambda}(\bm{k})c_{m,\lambda'}(-\bm{k})c_{m,\lambda''}(\bm{k}')c_{m,\lambda'''}(-\bm{k}') 12g\,,
\end{align}
where $g$ is the on-site anharmonicity parameter in the matter chain [Eq.~(\ref{eqn:H_M})]. Here, $c_{m,\lambda}(\bm{k})$ is the element of the $\lambda$ eigenvector of the dynamical matrix at $\bm{k}$ that corresponds to the matter atom.

\subsection{Vibrational dynamical mean-field theory}

\begin{figure*}[!ht]
    \centering
    \includegraphics[width=17cm]{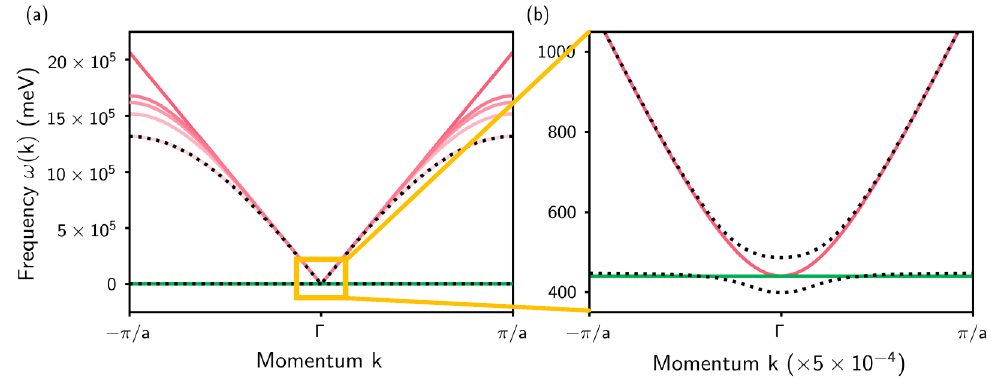}
    \caption{(a) The harmonic dispersion relation across the full Brillouin zone for the polariton lattice model given by Eq.~(\ref{eqn:H_polaritonLattice}) with $\omega_0 = \omega_m$ and $\eta=0.1$. The cavity and matter modes are shown in the pink and green solid lines, respectively, and the polariton states are shown in the black dotted lines. The analytical expression for the cavity dispersion is given by the darkest pink line, the dispersion for the cavity chain with nearest-neighbor interactions is given in the lightest pink, and dispersions for chains with longer-range  interactions are given by successively darker pink lines. (b) The harmonic dispersion relation in the region of the Brillouin zone around the $\Gamma$ point, which shows the energy region that is relevant for the polariton problem.}
    \label{fig2:harmonicDispersion}
\end{figure*}

This improved phonon basis from SCP can be used in combination with vibrational dynamical mean-field theory (VDMFT)~\cite{shih_anharmonic_2022} to compute the anharmonic lattice dynamics of the polariton system, including finite lifetimes and additional frequency shifts that are not captured by SCP.

In VDMFT, we calculate the anharmonic phonon Green's function (GF)~\cite{Cowley1963} of the periodic lattice,
\begin{equation}
    \bm{D}(\bm{k},\omega) = \int_0^\infty dt \frac{1}{k_B T}\langle \dot{\bm{u}}(\bm{k},t) \bm{u}^T(-\bm{k},0)\rangle\,,
\end{equation}
where $\langle \cdot \rangle$ denotes an equilibrium average at temperature $T$. This GF also satisfies a Dyson equation,
\begin{equation}
    \bm{D}^{-1}(\bm{k},\omega) = \omega^2\bm{1} - \bm{\Omega}^2(\bm{k}) - 2 \bm{\Omega}(\bm{k})\bm{\pi}(\bm{k},\omega)\,,
\end{equation}
where $\bm{\Omega}^2(\bm{k})$ is the dynamical matrix, including the mean-field contribution from SCP, and $\bm{\pi}(\bm{k},\omega)$ is the self-energy describing the additional, dynamical contributions to the anharmonicity. As there are two atoms in each unit cell, the GF, dynamical matrix, and self-energy are all $2\times 2$ matrices.

In DMFT, the dynamics of the periodic system are mapped onto those of a single unit cell (the ``system") coupled to a fictitious, self-consistently defined bath of harmonic oscillators with a tailored spectral density~\cite{GeorgesPhysRevB1992, GeorgesRevModPhys1996, kotliar2004strongly, KotliarRevModPhys2006, shih_anharmonic_2022, jasrasaria2024}. The local self-energy in the single unit cell, $\bm{\pi}(\omega)$, is computed through the solution of this so-called ``impurity problem," which is generally much simpler to solve than the periodic problem due to the small number of degrees of freedom in the system. This local self-energy is used to approximate the anharmonicity in the lattice GF, $\bm{\pi}(k, \omega)\approx\bm{\pi}(\omega)$; in this manner, the lattice GF and impurity problem are updated iteratively until self-consistency is achieved.

Our VDMFT approach provides a nonperturbative description of the local self-energy that describes frequency shifts, finite lifetimes, and mode-mixing due to anharmonicity. Through comparison to molecular dynamics (MD) simulations, which describe the exact dynamics for classical nuclei, VDMFT has been shown to be extremely accurate at a fraction of the cost~\cite{shih_anharmonic_2022, jasrasaria2024}. Furthermore, while MD requires the simulation of large supercells for high resolution of the Brillouin zone (BZ), the VDMFT GF is accessible at all points in the BZ. Finally, while this work focuses on computing the classical anharmonic dynamics, VDMFT can be used to treat anharmonic nuclear quantum effects~\cite{shih_anharmonic_2022}, which can only be described approximately using other techniques~\cite{ceriotti_colored-noise_2010, markland2018nuclear}. 

\section{Results\label{sec:results}}

\begin{figure*}[ht!]
    \centering
    \includegraphics[width=11.33cm]{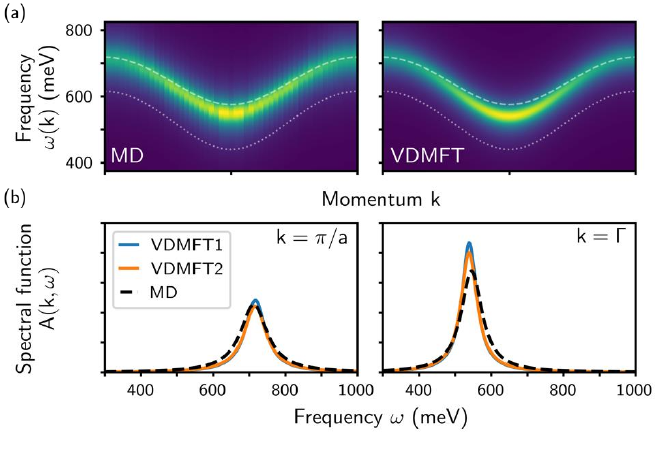}
    \caption{(a) Spectral functions at 300\,K for the isolated matter chain calculated using MD (left) and VDMFT (right). The dotted lines show the harmonic dispersion relation, whereas the dashed lines show the renormalized dispersion calculated using SCP at 300\,K. (b) Spectral functions at $k=\pi/a$ (left) and $k=\Gamma$ (right) calculated using MD and after the first and second iterations of VDMFT, indicating convergence of VDMFT and its excellent agreement with exact MD simulations.}
    \label{fig3:vdmftWater}
\end{figure*}

In this work, we motivate the choice of parameters in Eq.~(\ref{eqn:H_polaritonLattice}) using those of liquid water~\cite{TaoPNAS2020}. We set the lattice constant to $a=5.669$\,a.u. (3\,\AA, the water nearest-neighbor distance), the harmonic frequency to  $\omega_m=440$\,meV (3550\,cm$^{-1}$, the O-H bond stretch frequency), the nearest-neighbor interaction to $\Omega_m=215$\,meV (hydrogen bond energy), 
and the anharmonicity parameter to $g=4.3\omega_m^3$. As shown below, we use these parameters for the matter in combination with different cavity frequencies (at $k = 0$), $\omega_0$, and light-matter coupling strengths, $\eta$.

First, we calculate the harmonic dispersion relation of the polariton lattice model given by Eq.~(\ref{eqn:H_polaritonLattice}) with the cavity frequency tuned to be resonant to the harmonic frequency of the matter chain ($\omega_0 = \omega_m$) and a light-matter coupling strength of $\eta=0.1$. As illustrated in Fig.~\ref{fig2:harmonicDispersion}a, the noninteracting cavity and matter dispersions are shown in light pink and green, respectively, and the hybridized polariton dispersions are shown in the black dotted lines. Focusing on the dispersion of the noninteracting cavity mode, we see that it aligns well with the analytical expression for the dispersion of the cavity (given by the dark pink line, $\omega_0^{\text{an}}(k) = \sqrt{\omega_0^2 + c^2 k^2}$) at the center of the Brillouin zone (BZ) but underestimates this value at larger values of $k$. However, as discussed in Sec.~\ref{subsec:cavityHamiltonian}, this dispersion is systematically improvable through higher-order approximations of the second derivative that appears in Eq.~(\ref{eqn:2ndDerivative}), which would induce longer-range interactions between cavity atoms. Thus, a higher-accuracy description of the cavity modes requires a description that is more ``delocalized" in real-space, as well.

However, as shown in Fig.~\ref{fig2:harmonicDispersion}b, the BZ region of interest is that for which the cavity mode is near in energy to the matter mode, allowing for hybridization and the formation of the polariton bands. This energy range corresponds to a small region of the BZ that is very close to the $\Gamma$ point (i.e., center of the BZ). We see that in this region, the nearest-neighbor approximation made in Eq.~(\ref{eqn:H_polaritonLattice}) produces a cavity dispersion that coincides with the analytical expression. Additionally, in this region an energy splitting between the upper and lower polariton states appears, as expected.

Next, we include the effects of anharmonicity in our analysis. We perform SCP calculations of the isolated chain of matter atoms, and we find that the quartic anharmonicity in the matter chain causes a hardening of the harmonic frequency by 135\,meV at 300\,K, so that $\omega_{\text{m}}^{\text{SCP}}=\omega_{\text{m}}^0+135$\,meV. We then use the quasiparticle basis obtained using SCP in combination with VDMFT to compute the anharmonic GF and spectral function of the matter chain, $\bm{A}(\bm{k},\omega) = -\pi^{-1}\text{Tr}[\Im \bm{D}(\bm{k},\omega)]$, following the approach described in detail in our previous work~\cite{jasrasaria2024}. As shown in Fig.~\ref{fig3:vdmftWater}b, we see convergence of the anharmonic spectral function in one iteration, and we see excellent agreement with the exact anharmonic spectral function computed using MD across the entire BZ. Our SCP+VDMFT approach captures frequency renormalization caused by anharmonicity, predicting a slightly smaller shift of $\omega_{\text{m}}^{\text{VDMFT}}=\omega_{\text{m}}^0+110$\,meV, as well as finite lifetimes due to phonon-phonon scattering. The finite lifetimes manifest as the broad linewidth in the anharmonic spectral function, which here corresponds to lifetimes of ~100\,fs and are in agreement with experimental measurements of the lifetime of the O-H stretch vibration in bulk water~\cite{kraemer2008temperature}. 

As the chain of cavity atoms is purely harmonic and the light-matter coupling is relatively weak, we assume that it does not affect the self-energy of the isolated chain of matter atoms (i.e., that the light-matter coupling does not affect anharmonicity beyond the static mean-field level). Thus, to account for anharmonicity in the polariton model, we perform SCP calculations of the coupled light-matter system, and we use that quasiparticle basis along with the self-energy calculated for the isolated matter chain to compute the anharmonic GF and spectral function.

\begin{figure*}[!ht]
    \centering
    \includegraphics[width=17cm]{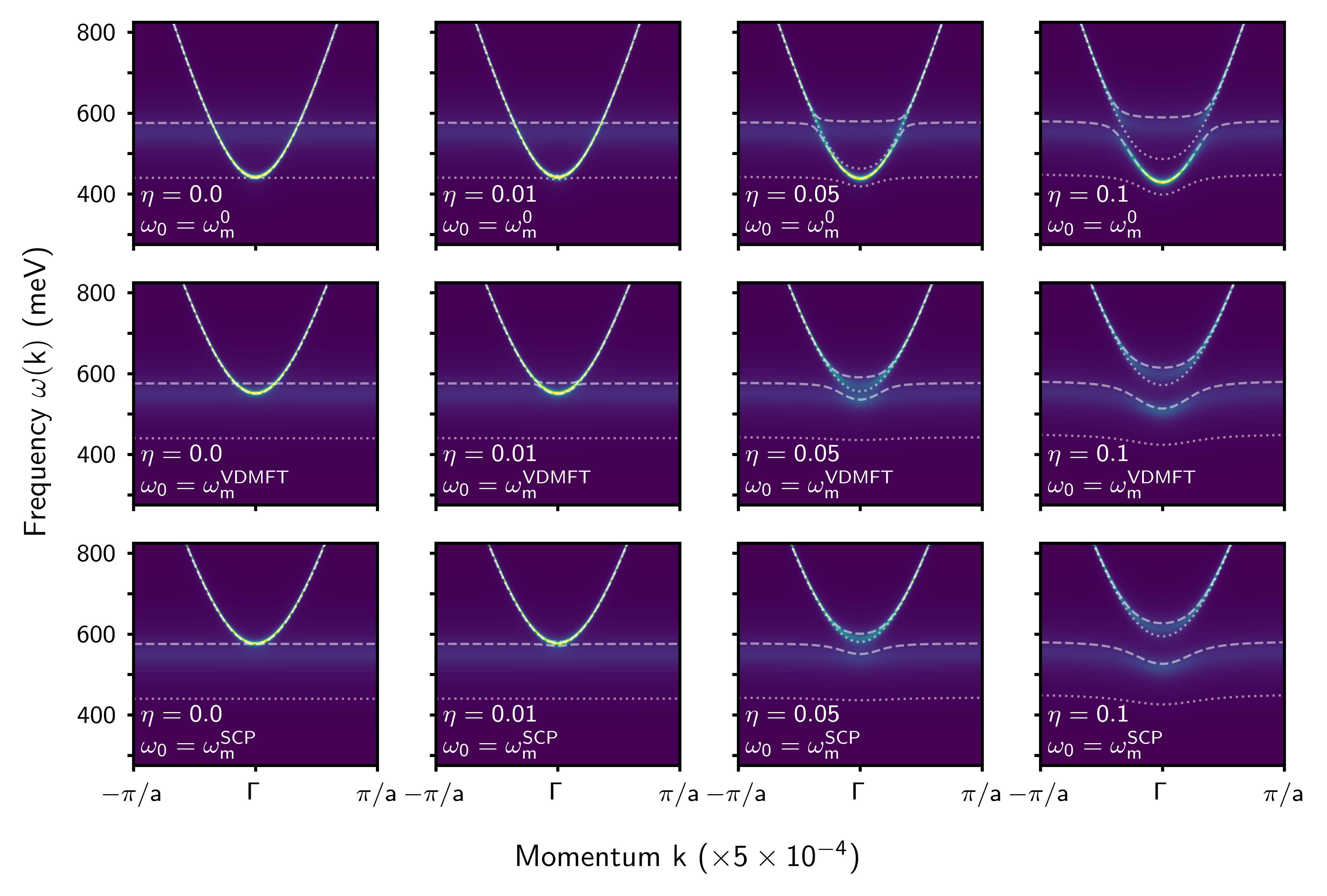}
    \caption{Spectral functions for the polariton lattice model at different light-matter coupling strengths and cavity frequencies. The dotted lines show the harmonic dispersion relation, the dashed lines show the renormalized dispersion calculated using SCP at 300\,K, and the colored heat maps show the anharmonic spectral functions calculated using SCP+VDMFT at 300\,K.}
    \label{fig4:spectralFunction}
\end{figure*}

Figure~\ref{fig4:spectralFunction} shows the harmonic and anharmonic dispersion relations of the polariton model for a variety of cavity frequencies, $\omega_0$, and light-matter coupling strengths, $\eta$. The harmonic dispersion relation is illustrated in the dotted lines, the renormalized SCP dispersion is shown in the dashed lines, and the full, anharmonic spectral function calculated using VDMFT is illustrated by the color maps. When the cavity is tuned to be resonant to the harmonic frequency of the matter chain, $\omega_0 = \omega_{\text{m}}^0$ (top row of Fig.~\ref{fig4:spectralFunction}), the cavity and the matter bands cross away from the $\Gamma$ point due to the anharmonic frequency renormalization of the matter band. Interestingly, as $\eta$ increases, the lower polariton state retains the cavity's narrow linewidth (i.e., long lifetime) near $k=\Gamma$ and broadens away from the $\Gamma$ point as it gains more matter character. The converse is true for the upper polariton state. This result demonstrates how cavities can be used to form polariton states with tailored lifetimes (or tailored levels of anharmonicity), in addition to tailored energies and dispersion relations.

When the cavity is tuned to be resonant to the VDMFT frequency, $\omega_0 = \omega_{\text{m}}^{\text{VDMFT}}$ (middle row of Fig.~\ref{fig4:spectralFunction}), the Rabi splitting occurs at the $\Gamma$ point, as expected, and the lower and upper polariton states have similar linewidths. Again, as the light-matter coupling decreases away from $k=\Gamma$, the polariton states take on linewidths that reflect whether they are primarily of light or matter character. Similar behavior is observed when the cavity is tuned to be resonant to the SCP frequency, $\omega_0 = \omega_{\text{m}}^{\text{SCP}}$ (bottom row of Fig.~\ref{fig4:spectralFunction}).

\begin{figure*}[!ht]
    \centering
    \includegraphics[width=17cm]{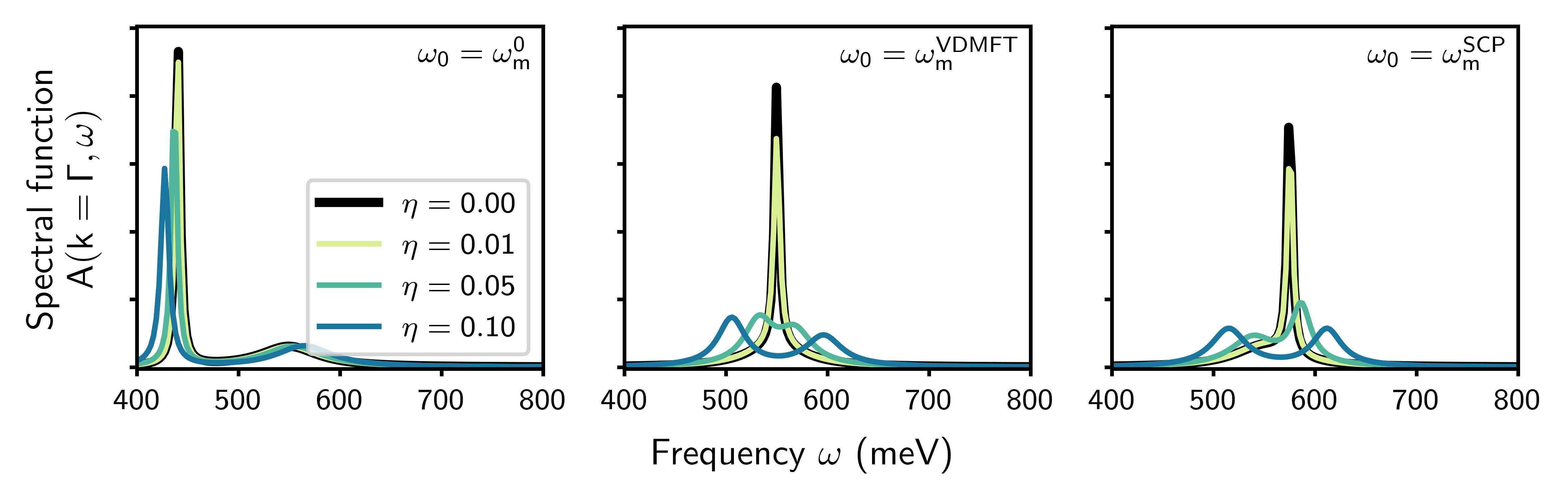}
    \caption{The anharmonic spectral function at $k=\Gamma$ calculated using SCP+VDMFT at 300\,K for different light-matter coupling strengths and cavity frequencies.}
    \label{fig5:spectralCut}
\end{figure*}

To better understand the dependence of the polariton anharmonic spectral functions on the light-matter coupling, we illustrate the spectral functions at the $\Gamma$ point, $A(k=\Gamma,\omega)$, in Fig.~\ref{fig5:spectralCut}. As discussed above, when $\omega_0 = \omega_{\text{m}}^0$, the energy gap between upper and lower polariton bands, as well as the linewidth of the lower polariton bands, increases slightly with increasing $\eta$, but there is minimal effect at $k=\Gamma$ as the cavity is essentially tuned to be off-resonant with the matter chain. When the cavity is tuned to be resonant with the frequency that maximizes the matter anharmonic spectral function, $\omega_0=\omega_{\text{m}}^{\text{VDMFT}}$, the spectral function at $k=\Gamma$ remains symmetric as $\eta$ increases, both in terms of the energy shifts and linewidths of the polariton states. The polariton spectral function shows similar behavior at $\omega_0=\omega_{\text{m}}^{\text{SCP}}$, especially for larger $\eta$ values.

\begin{figure*}[!ht]
    \centering
    \includegraphics[width=17cm]{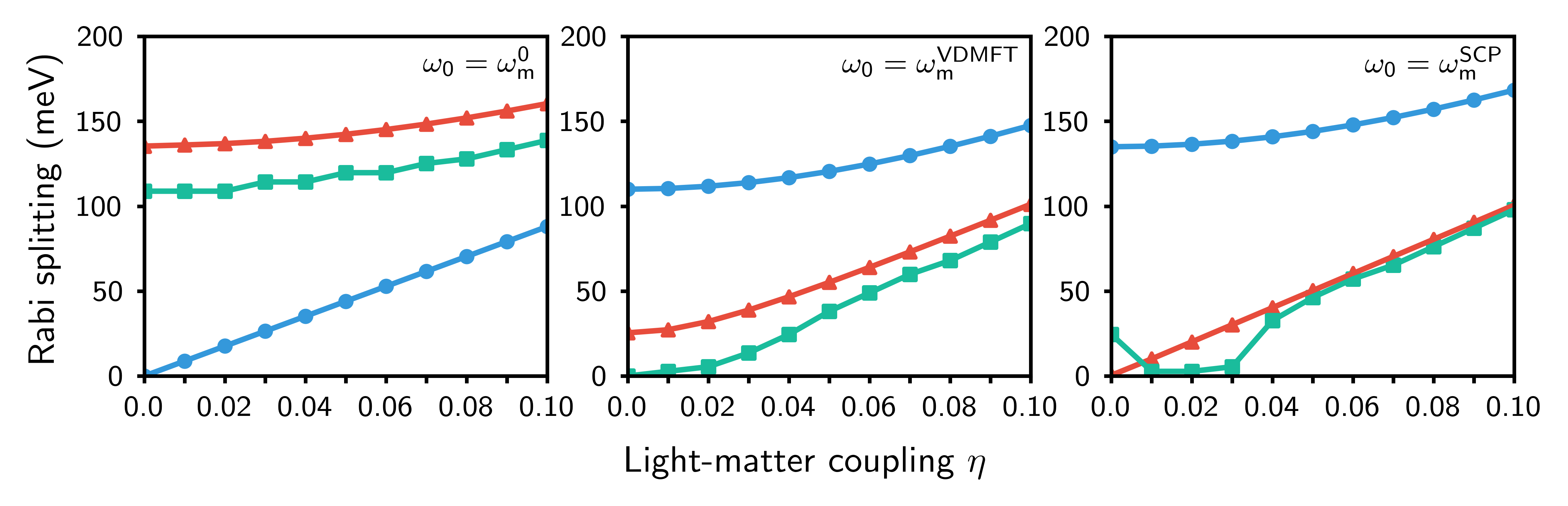}
    \caption{The Rabi splitting as a function of light-matter coupling calculated from the harmonic dispersion (blue circles), SCP dispersion (red triangles), and SCP+VDMFT spectral function (turquoise squares).}
    \label{fig6:rabiSplitting}
\end{figure*}

Finally, we consider the Rabi splitting, which we define as the difference between the upper and lower polariton state frequencies at different levels of theory, $\Omega_{\text{R}} = \omega_{\text{UP}}(k=\Gamma) - \omega_{\text{LP}}(k=\Gamma)$. As illustrated in the left panel of Fig.~\ref{fig6:rabiSplitting}, when $\omega_0 = \omega_{\text{m}}^0$, the Rabi splitting $\Omega_{\text{R}}^0=0$ in the noninteracting limit ($\eta=0$) and increases linearly with $\eta$, as expected. However, $\Omega_{\text{R}}^{\text{SCP}}$ and $\Omega_{\text{R}}^{\text{VDMFT}}$ have large values in the noninteracting limit due to the anharmonic frequency renormalization and increases only slightly with stronger light-matter coupling.

Interestingly, the middle panel of Fig.~\ref{fig6:rabiSplitting} shows that for $\omega_0 = \omega_{\text{m}}^{\text{VDMFT}}$, the Rabi splitting $\Omega_{\text{R}}^{\text{VDMFT}}$ increases very slightly with light-matter coupling for small $\eta$ values. This is because of the significant spectral overlap between the broad matter band and the narrow cavity mode. As $\eta > 0.03$, the Rabi splitting increases significantly with linear behavior. This result demonstrates that the expected linear dependence of the Rabi splitting on light-matter coupling strength does not hold, even for weak coupling, when anharmonicity is included. This is reminiscent of Fano resonance phenomena, when coupling between a discrete and a continuum state leads to unique lineshapes of the scattered states. 

Finally, when $\omega_0 = \omega_{\text{m}}^{\text{SCP}}$, the right panel of Fig.~\ref{fig6:rabiSplitting} shows that $\Omega_{\text{R}}^{\text{SCP}}$ increases linearly with $\eta$, analogous to $\Omega_{\text{R}}^0$ for when $\omega_0 = \omega_{\text{m}}^0$, which makes sense as SCP is an effective harmonic theory. However, $\Omega_{\text{R}}^{\text{VDMFT}}$ has non-monotonic behavior with increasing $\eta$. For small light-matter coupling strengths, the Rabi splitting becomes negligibly small as the cavity and matter bands are significantly mixed due to their spectral overlap. However, for $\eta > 0.05$, the $\Omega_{\text{R}}^{\text{VDMFT}}$ and $\Omega_{\text{R}}^{\text{SCP}}$ values coincide.
Clearly, including anharmonic effects, both for frequency renormalization and broadening, are important to an accurate analysis of polariton dispersions.

\section{Conclusions\label{sec:conclusions}}

In conclusion, we have presented a new theoretical framework for modeling vibrational polariton systems beyond the long wavelength approximation, which is important to understanding a variety of experimentally-observed phenomena. We performed a simple transform to the standard dipole gauge light-matter Hamiltonian, which includes many molecules coupled to several $k$-dependent cavity modes, to show that it can be modeled as two coupled periodic lattices in real space, a molecular lattice coupled to a lattice of spatially localized radiation modes. This Hamiltonian provides an alternative physical picture for light-matter coupling, and provides new insight regarding the scaling of light-matter coupling with the number of molecules inside the cavity. This localized framework may be useful for understanding how delocalized, collective light-matter coupling results in changes of local chemical reactivities, and will be the subject of future work.

Within this framework, we show that VDMFT is a simple tool for calculating momentum-resolved spectra of the polariton system. VDMFT includes a nonperturbative description of anharmonicity, and it is both accurate and efficient, especially when considering sampling of the BZ in the energy region that is relevant for the polariton problem. Furthermore, nuclear quantum dynamics can be straightforwardly incorporated into the VDMFT framework~\cite{shih_anharmonic_2022} to understand their effects on polariton spectra. 

Through the application of VDMFT to a simple molecular model, we demonstrate that inclusion of anharmonicity in the matter degrees of freedom significantly affects vibrational polariton states and their spectra. Considering temperature-dependent frequency renormalization due to anharmonicity strongly alters the frequency at which to tune the cavity to be resonant. Beyond frequency renormalization, anharmonicity can impart broad linewidths on the spectral functions of both the matter and cavity degrees of freedom due to finite lifetimes caused by phonon-phonon scattering. We show that these linewidths can be tuned via coupling to harmonic cavity modes, which would have implications on a variety of other observables, including thermal transport properties, the calculation of which is the subject of future work.

Additionally, spectral overlap between narrow cavity and broad molecular states affects the Rabi splitting between hybridized polariton states and its behavior with increasing light-matter coupling strengths, which deviates from the linear dependence on light-matter coupling strength that is predicted by harmonic analysis. Thus, accurately simulating anharmonicity can affect the design of polariton states for optimized cavity-modified properties.

\textit{Acknowledgements.}
D.J. and T.C.B were supported by the US Air Force Office of Scientific Research under Grant No. FA9550-21-1-0400.
A.M. and D.R.R. were supported by NSF CHE-2245592.
The Flatiron Institute is a division of the Simons Foundation.

%

\end{document}